\documentclass[apl,reprint,graphicx]{revtex4-1}
\usepackage{siunitx}
\usepackage{graphicx}

\draft % marks overfull lines with a black rule on the right

\begin{document}

%\preprint{}

\title{Individually addressable double quantum dots formed with nanowire polytypes and identified by epitaxial markers} %Title of paper



\author{D. Barker}
\email[E-mail: ]{david.barker@ftf.lth.se}
%\thanks{}
\author{S. Lehmann}
\author{L. Namazi}
\author{M. Nilsson}
\author{C. Thelander}
\author{K. A. Dick}
\altaffiliation[Also at ]{Centre for Analysis and Synthesis, Lund University, Box 124, 22100 Lund, Sweden}
\affiliation{NanoLund and Solid State Physics, Lund University, Box 118, 22100 Lund, Sweden}
\author{V. F. Maisi}
\email[E-mail: ]{ville.maisi@ftf.lth.se}
\affiliation{NanoLund and Solid State Physics, Lund University, Box 118, 22100 Lund, Sweden}


\date{\today}

\begin{abstract}
\noindent Double quantum dots (DQDs) hold great promise as building blocks for quantum technology as they allow for two electronic states to coherently couple. Defining QDs with materials rather than using electrostatic gating allows for QDs with a hard-wall confinement potential and more robust charge and spin states. An unresolved problem is how to individually address these quantum dots, which is necessary for controlling quantum states. We here report the fabrication of double quantum dot devices defined by the conduction band edge offset at the interface of the wurtzite and zinc blende crystal phases of InAs in nanowires. By using sacrifical epitaxial GaSb markers selectively forming on one crystal phase, we are able to precisely align gate electrodes allowing us to probe and control each QD independently. We hence observe textbook-like charge stability diagrams, a discrete energy spectrum and electron numbers consistent with theoretical estimates and investigate the tunability of the devices, finding that changing the electron number can be used to tune the tunnel barrier as expected by simple band diagram arguments.
\end{abstract}


\maketitle %\maketitle must follow title, authors, abstract and \pacs

\begin{figure}[t]
\begin{centering}
\includegraphics{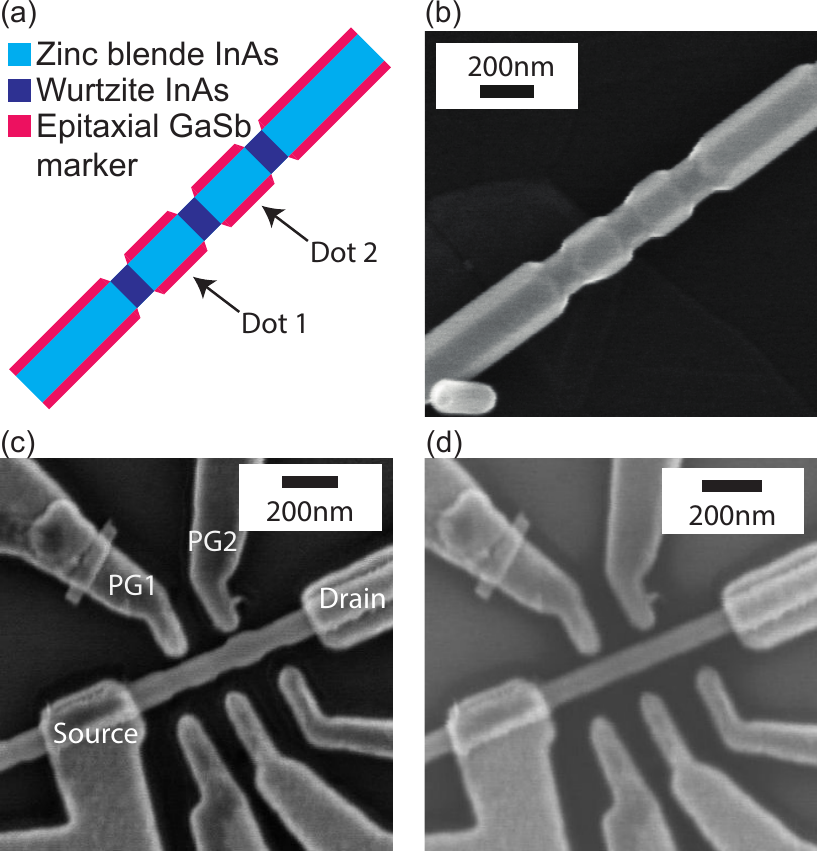}
  \caption{(a) Cross-sectional sketch of  a nanowire, showing the different polytypes as well as the sacrificial epitaxial markers. (b) SEM image of an InAs nanowire with GaSb shell grown preferentially on zinc blende InAs. (c) SEM image of Device B before etching of the epitaxial markers showing alignment of plunger gates with each dot.  (d) The same device as in (c) after etching of GaSb.}
  \label{fgr:semimages}
\end{centering}
\end{figure}

\begin{figure}[t]
\begin{centering}
\includegraphics[width=\columnwidth]{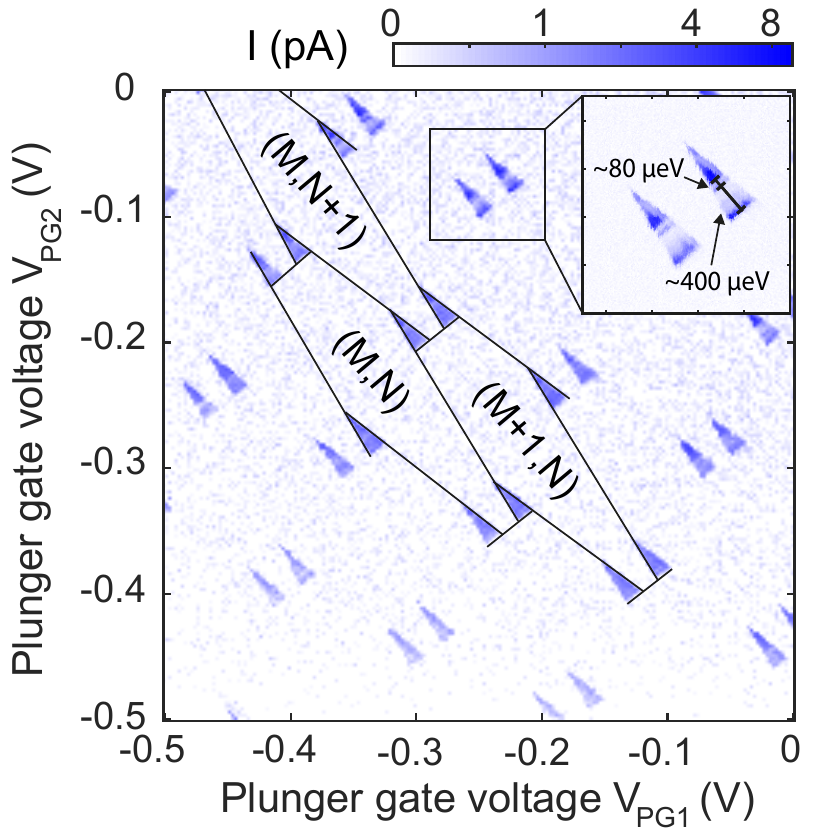}
  \caption{Charge stability diagram for Device A at low plunger gate voltages with the back gate voltage set to $V_{BG} = \SI{1}{\volt}$. N and M are electron occupations of the two dots. Inset: High resolution measurements from a pair of finite bias triangles. Excited states can be identified from resonance lines inside the triangles. Along the height of the triangle, there are two states with \SI{400}{\micro\electronvolt} and \SI{80}{\micro\electronvolt} spacings.}
  \label{fgr:closedregime}
\end{centering}
\end{figure}

\begin{figure}[t]
\begin{centering}
\includegraphics{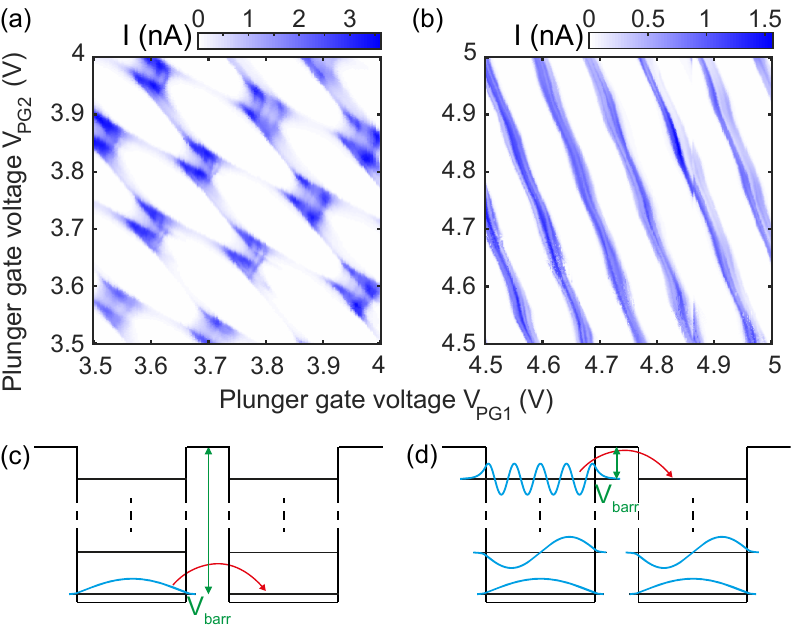}
  \caption{(a-b) Charge stability diagrams for Device A at higher plunger gate voltages. The back gate voltage is still $V_{BG} = \SI{1}{\volt}$. (c-d)Band diagrams for low (c) and high (d) electron numbers and plunger gate voltages. The green arrow, representing the effective barrier height $V_{barr}$ is smaller when more states are filled. The high energy wave functions penetrating the barrier further are also illustrated.}
  \label{fgr:higherregimes}
\end{centering}
\end{figure}

\noindent When electrons are spatially confined in semiconductor quantum dots (QDs), they form bound states with discrete energy levels. These systems have drawn much attention both experimentally and theoretically~\cite{Wiel2003, Hanson2007} as they form atom-like structures in solid-state. One system of particular importance is the double quantum dot (DQD) where two discrete electronic states couple coherently, making it the building block of charge and spin qubits~\cite{Wiel2003, Hanson2007, Loss1998, Landig2017, Zwanenburg2013, Mi2017}. QDs are also elemental in other semiconductor quantum systems that are promising for use in quantum computers and quantum systems in general~\cite{DiVincenzo1998, Landig2017, Zwanenburg2013}, such as Majorana fermions~\cite{Lutchyn2010, Oreg2010, Mourik2012, Albrecht2016}. QDs are commonly defined by gate depletion~\cite{Hanson2007,  Zwanenburg2013, Fasth2005}, but progress has been made in material-defined QDs as well~\cite{Tarucha1996, Bjork2004, Nilsson2016}. The material-defined approach allows for more well-defined features and less coupling to external noise. In this letter, we utilize recently developed InAs polytype bandgap engineering~\cite{Lehmann2013, Nilsson2016, Nilsson2017} to define DQDs with a hard-wall potential. With epitaxial markers, we gain control of the individual dots and, demonstrate the honeycomb-shaped charge stability diagrams of material-defined DQDs and the robustness of the system with a wide range of electron populations.\\
\\The most common approach to forming quantum dots for transport experiments is to start from a material system that is already structurally confined in one or two dimensions and use electrostatic gating to confine the remaining dimensions. Examples here include two-dimensional electron gases~\cite{Hanson2007, Zwanenburg2013}, one-dimensional carbon nanotubes and semiconductor nanowires~\cite{Fasth2005}. These partially gate-defined QDs have a smoothly changing confinement potential and hence their size and form are not very well-defined. In addition, any noise that couples to the gate electrodes defining the dots will change the shape and polarization of the dot which can lead to decoherence. QDs defined by materials have sharper confinement potentials, which means that the dot size and shape, and therefore also the energy levels, are less susceptible to noise. The first dots made this way used a double barrier semiconductor heterostructure which was then patterned into mesas to define disc-shaped dots~\cite{Tarucha1996}. More recently, single and few donor sites have been utilized to obtain confinement potentials similar to the atoms~\cite{Zwanenburg2013, Fuechsle2010, Tan2010}. In nanowires, QDs have been made both by introducing segments of another semiconductor with a larger bandgap~\cite{Bjork2004} as well as introducing segments of another crystal phase~\cite{Nilsson2016} during growth. Thanks to the materials, such QDs can have very high confinement energies and reveal an electron shell filling pattern expected of a cylindrically symmetric system~\cite{Tarucha1996, Bjork2004}.\\
\\Extending the material defined single dots to material defined DQDs is very attractive to obtain the coupling of discrete energy levels.  However, individual addressing of quantum states and dots in material-defined DQDs is not straightforward, and has therefore severely limited their use. In previous reports of serial heterostructure DQDs, a common gate was used to control the potential in both dots~\cite{Ono2002, Fuhrer2007}. This was partly due to difficulties in resolving the structure during the fabrication process, and partly that the QD pitch is too small for standard lithography processes. In this work we make use of sacrificial epitaxial markers of GaSb that selectively grow on one crystal phase (zinc blende, used for QD and leads), and is suppressed on the other (wurtzite, used for tunnel barriers), as presented in Figure~\ref{fgr:semimages}(a). The conduction band edge offset is also much shallower than in other strucurally defined QDs, which allows for a much larger QD separation, and therefore individual addressing of quantum dots and barriers by local gates. Our long tunnel barriers, up to several tens of nm instead of the more typical case of a few nm~\cite{Tarucha1996, Bjork2004, Simmons1963, Aref2014, Zeng2015} are also appealing to construct new device functionalities such as spin rotations across the barriers via spin-orbit interaction.\\
\\In order to obtain the material-defined DQDs, we used a nanowire growth process developed in Ref. \citenum{Lehmann2013}. This process allows for sharp crystal phase boundaries in nanowires and has previously been used to grow single dots of high quality~\cite{Nilsson2016,Nilsson2017}. The nanowire growth was controlled so that three regions of wurtzite (WZ) InAs are spaced out among zinc blende (ZB) structure, as shown in Figure~\ref{fgr:semimages}(a). The DQD is then defined in the ZB by the tunnel barriers created by WZ due to the conduction band edge offset at the interface of the two polytypes, likely in combination with a difference in surface charge between the materials~\cite{Dayeh2009}. The epitaxial markers enabling the alignment of electrodes for transport experiments were formed by a \SI{20}{\nano\meter} thick GaSb shell which was grown selectively on the ZB InAs using the process described in Ref \citenum{Namazi2015}. Figure~\ref{fgr:semimages}(b) presents an SEM micrograph of the section of interest of such a wire after the growth where the attained structure is clearly visible.\\
\\To investigate if  this new approach can result in high quality DQDs suitable for building quantum devices, we fabricated gate electrodes and source-drain contacts and performed transport experiments. The wires were first transferred from the growth chip to a highly doped Si wafer with a \SI{200}{\nano\meter} thermal oxide. The substrate could then be used as a global back gate. After transfer, the electrodes and contacts were fabricated with electron beam lithography. Thanks to the selectively grown GaSb shell, these could be aligned using SEM imaging. The alignment to the QDs can be seen in Figure~\ref{fgr:semimages}(c). After the positioning, the GaSb was selectively etched by submerging the chip in MF-319 developer for three minutes~\cite{Gatzke1997}. The etching process was performed either before (Device A)  or after (Device B) deposition of \SI{25}{\nano\meter}/\SI{75}{\nano\meter} Ni/Au contacts with both methods giving good results. Before metallization, resist residues are removed by a \SI{20}{\second} oxygen plasma and native oxide is etched by submerging the sample in 10:1 buffered oxide etch for \SI{5}{\second}. The final Device B is shown in Figure~\ref{fgr:semimages}(d).\\
\\Electrical measurements were then performed in a dilution refrigerator at an electron temperature $T\approx\SI{50}{\milli\kelvin}$ with the following procedure: A \SI{1}{\milli\volt} bias voltage was applied between source and drain and the back gate voltage was increased until the wire became conducting with a roughly \SI{10}{\pico\ampere} current. Back gate voltages were $V_{BG} = \SI{1}{\volt}$ for Device A and $V_{BG} = \SI{0}{\volt}$ for Device B. After that, the gates labelled "PG1" and "PG2" in Figure~\ref{fgr:semimages}(c) were used as plunger gates to control the electron population of each dot. These were swept around zero voltage as we monitored the source-drain current. This relatively straightforward procedure highlights another advantage of the material-defined systems, namely the simplicity of operation. In a gate-defined system, changing one gate voltage will affect all energy levels and all tunnel barriers to a larger degree, so more work will be needed to tune the system to a suitable operation point. Figure~\ref{fgr:closedregime}(a) shows the results of these measurements. We observe the typical hexagonal DQD charge stability pattern where transport is blocked except in triangle-shaped regions known as finite bias triangles. Connecting the triangles yields a honeycomb pattern where each hexagon represents a charge state~\cite{Wiel2003}. The honeycomb pattern being present without requiring any tuning other than making the system conducting with a global back gate proves the polytype structure defines the DQD as intended.\\
\\ To further characterize the device and confirm the connection between dimensions defined by the growth and the device parameter values, we determined from the charge stability diagrams the charging energies $E_{Cj} = e^2/C_j$, the interdot coupling energy $E_{Cm} = (e^2/C_m)((C_1C_2/C_{m}^2)-1)^{-1}$  and the lever arms $\alpha_{ij}$ by following Refs \citenum{Wiel2003} and \citenum{Taubert2011}. The lever arm is a measure of the coupling between plunger gate $i$ and dot $j$. $C_{j}$ is the total capacitance to island $j$ and $C_m$ is the capacitance between the dots. These parameters were extracted for two different devices. Device A (GaSb etched before metallization) had dot lengths of about \SI{80}{\nano\meter} and \SI{100}{\nano\meter} with 50-\SI{60}{\nano\meter} barriers. Device B had both dots with a length of around \SI{110}{\nano\meter} and 50-\SI{60}{\nano\meter} barriers. The results are collected  in Table~\ref{tbl:numbers}. The charging energies $E_{C1,2}$ are the same in Device B with similar dot sizes, whereas they differ for Device A with dissimilar dot sizes. The charging energies are inversely proportional to the length, as expected for charging energy set by either self-capacitance or capacitance towards back gate. This further confirmed that the dots are formed by the material, as the charging energies are directly linked to the ZB segment size. The lever arms are significantly higher for Device B because the gates were placed closer to the wire than for Device A (\SI{25}{\nano\meter} compared to \SI{80}{\nano\meter}).\\
\begin{table}
\begin{centering}
  \caption{Charging energies and lever arms for two DQD devices extracted from charge stability diagrams.}
  \label{tbl:numbers}
  \resizebox{\columnwidth}{!}{%
  \begin{tabular}{llll}
    \hline
     & Device A  & Device A  & Device B  \\
     & (Figure~\ref{fgr:closedregime}) & (Figure~\ref{fgr:higherregimes}(a)) & \\
    \hline
    $E_{C1}$ (meV) & $5.8$ & 5.3  & $4.2$   \\
    $E_{C2}$ (meV) & $6.7$ & 4.9 & $4.2$ \\
    $E_{Cm}$ (meV) & $1.8$ & 1.2 & $1.5$  \\
    \hline
    $\alpha_{11} (eV/V)$ & $0.050$ & 0.042 & $0.217$ \\
    $\alpha_{12} (eV/V)$ & $0.031$ & 0.026 & $0.125$\\
    $\alpha_{21} (eV/V)$ & $0.027$ & 0.021 & $0.077$\\
    $\alpha_{22} (eV/V)$ & $0.044$ & 0.033 & $0.202$\\
    \hline
  \end{tabular}
}
\end{centering}
\end{table}
\\When investigating the finite bias triangles in more detail, we observe excited states such as the ones in the inset of Figure~\ref{fgr:closedregime}. These features would not be present if there was a continuum of states, so their existence is evidence that the WZ barriers are strongly confining the electrons. From Figure~\ref{fgr:closedregime} we determined the level spacing to be around \SI{400}{\micro\electronvolt} and \SI{80}{\micro\electronvolt} for those excited states. These were typical values for the device, with most level spacings being in the 300-\SI{400}{\micro\electronvolt} range. Approximating the QDs as cylinders with \SI{80}{\nano\meter} length and diameter and sharp and high confinement potential yields level spacings ranging from some tens of $\mu$eV to a few meV consistent with the range observed experimentally in our device. Here we used the effective electron mass $m^{*} = 0.023m_e$ of InAs, where $m_e$  is the electron rest mass.\\
\\One of the main advantages of the gate-defined QDs is that the electron numbers and tunnel couplings can be tuned almost independently~\cite{Fasth2005, Petta2004}. With material defined dots, this flexibility is expected to be lost to at least some degree. To investigate the trade-off between tunability and rigidity of the confining potential, we measured the stability diagrams at higher plunger gate voltages as presented in Figure~\ref{fgr:higherregimes}. In panel (a) we increased the plunger gates from \SI{0}{\volt} to \SI{4}{\volt}, increasing the electron number by approximately 20-30 in each dot. At the same time, we observe that the overall current through the DQD increases by more than two orders of magnitude, which indicates stronger tunnel coupling. Additionally, we obtain current between the finite bias triangles. These lines arise from cotunneling, which also gets more probable the stronger the tunnel couplings are. We explain this finding with the help of the band diagrams illustrated in Figure~\ref{fgr:higherregimes}(c-d). As we go to higher gate voltages, the dots are populated with more electrons residing at higher energy states. For the high energy states, the effective tunnel barrier $V_{barr}$ is smaller and tunnel couplings larger in line with the observations. In order to obtain full tunability, we tried to adjust the tunnel barrier heights with the barrier gates shown in Figure~\ref{fgr:semimages} and the back gate, which resulted in only shifts in the stability diagrams but no change in the couplings. Therefore, we still maintain the tuning of either electron population or the tunnel coupling but cannot change both of them independently, which reflects the rigidity of our confining potential. However, by changing the thickness of the tunnel barriers and/or the quantum dots, we expect to reach any given electron number and tunnel coupling configuration.\\
\\At very large electron numbers, the wavefunctions will penetrate deeper into the barrier, also increasing the coupling and even modifying the shape of the dot. Figure~\ref{fgr:higherregimes}(b) presents an extreme case of this scenario where we increased the electron number even further by approximately ten. Single dot features are observed with transport lines not matching any of the DQD case, meaning the two QDs have merged into one due to strong gate-induced deformation. By considering the number of electronic states in the cylindrical approximation residing below the \SI{135}{\milli\electronvolt} WZ-ZB conduction band edge offset reported in Ref~\citenum{Chen2017}, we estimate to have 30-40 bound states available in a dot of this size. Hence, when changing the electron numbers from Figure~\ref{fgr:closedregime} at $V_{PGi} = \SI{0}{\volt}$ to Figure~\ref{fgr:higherregimes}(a), we add a considerable fraction of the total number of electrons fitting in the well without destroying the DQD pattern or changing the parameter values of Table~\ref{tbl:numbers} considerably, even when getting close to the breakdown occurring at $V_{PGi} = \SI{5}{\volt}$ in Figure~\ref{fgr:higherregimes}(b). The high electron numbers at $V_{PGi} = \SI{4}{\volt}$ results in lower charging energies compared to the low electron number case (see Table~\ref{tbl:numbers}) when $V_{PGi} = \SI{0}{\volt}$, which is in line with the wavefunctions penetrating deeper into the barrier and hence increasing the junction capacitances. The lever arms reducing also fits with this picture, as the capacitance ratios that define the lever arms are lowered. Based on the estimation of bound states and the number of electrons added, we find that the QDs hold a few tens of electrons at low plunger gate voltages.\\
\\In conclusion, we used sacrificial epitaxial markers to align gate electrodes to fully material-defined DQDs with sharp confinement potential and well-defined and known dimensions. This allowed us to  independently tune the electron population of each dot. The device structure is more predictable and expected to be less susceptible to external noise than corresponding structures fabricated with other methods. We anticipate them to be important to studies of quantum phenomena and in particular contribute to the trend of increasing coherence times for DQD qubits~\cite{Landig2017, Mi2017, Stockklauser2017}. We found that the number of electrons that can fit into the dots is consistent with theoretical estimates. In addition, we also showed the relation between the dot size and the resulting charging energy and studied the tunability of the charge numbers and tunnel couplings, finding the devices to be tunable in either electron number or couplings but not both.\\
\\We acknowledge financial support by NanoLund and the Knut and Alice Wallenberg Foundation (KAW) via Project No. 2016.0089 and Wallenberg Centre for Quantum Technology.

% If in two-column mode, this environment will change to single-column format so that long equations can be displayed. 
%\begin{widetext}
%$$\mbox{put long equation here}$$
%\end{widetext}

% Figures should be put into the text as floats. 
% Use the graphics or graphicx packages (distributed with LaTeX2e).
% Put the label that you will use with \ref{} command in the braces of the \label{} command.
%
% \begin{figure}
% \end{figure}

% \begin{table}
% \end{tabular}
% \end{table}

%\begin{acknowledgments}
% Put your acknowledgments here.
%\end{acknowledgments}

\bibliography{bibliography}

\end{document}